\documentclass[pra,aps]{revtex4}

\usepackage{amsmath}
\usepackage{ulem}
\usepackage{bm}
\usepackage{graphics}
\usepackage[dvips]{color}
\usepackage{dcolumn}

\begin{document}
\title{Population Dynamics of a Spin-1 Bose Gas Above the Bose-Einstein Transition Temperature}
\author{Yuki Endo and Tetsuro Nikuni}
\affiliation{Department of Physics, Faculty of Science, Tokyo University of Science, 
1-3 Kagurazaka, Shinjuku-ku, Tokyo, Japan, 162-8601}
\date{\today}

\begin{abstract}
We study population dynamics of a trapped spin-1 Bose gas above the Bose-Einstein transition temperature. Starting from the semiclassical kinetic equation for a spin-1 gas, we derive a coupled rate equations for the populations of internal states. Solving the rate equations, we discuss the dynamical evolution of spin populations. We also estimate the characteristic timescale in which the system reaches equilibrium. Finally, we briefly discuss how the presence of the condensate will affect the population dynamics.
\end{abstract}

\maketitle
\section{Introduction} 
Bose-Einstein condensation (BEC) in dilute atomic gases have been extensively studied both theoretically and experimentally since its experimental realization in 1995~\cite{Anderson1995,Davis1995}. In particular, BEC involving internal degree of freedom has been extensively studied. The first experiment confining BEC with multi internal states was achieved by JILA group~\cite{Myatt1997}. Shortly afterward, they succeeded in confining ${\rm ^{87}Rb}$ atoms with two internal states $\left| F=2,m_F =1 \right\rangle$ and $\left| F=1,m_F =-1 \right\rangle$, which is called spin-1/2 Bose gas~\cite{Hall1998}. Spin-1/2 Bose gases are known to exhibit the collective spin dynamics due to the exchange effect even above the Bose-Einstein transition temperature (${T_{\rm BEC}}$)~\cite{Lewandowski2002,McGuirk2002,Oktel2002_2,Fuchs2002,Williams2002,Nikuni2002,Nikuni2003}.

On the other hand, the spinor condensate trapped in an optical trap has been achieved experimentally. An optical trap can trap atoms with maintaining spin degree of freedom in contrast to magnetic trap where the spin degree of freedom is frozen. The MIT group first succeeded in creating a spinor condensate in an optical trap for ${\rm ^{23}Na}$ (so called spin-1 BEC)~\cite{Stenger1998}. In the spin-1 system, the spin-spin interaction that describes the collision of two atoms exchanging there internal states plays an important role, even though they are very small compared with the usual spin-independent interaction~\cite{Stenger1998,Kurn1998,Law1998,Miesner1999,Zhang2003,Chang2004,Schmaljohann2004,Erhard2004,Higbie2005,Sadler2006}.

Moreover in the spin-1 system, these components couple automatically through intrinsic spin-spin interaction. This is in contrast to the spin-1/2 system, where these components couple with each other through the exchange effect. We thus expect the spin-1 Bose gas to exhibit much richer  spin dynamics than in the spin-1/2 system.

Even though most of the studies in the dynamics of spin-1 Bose gases have focused on a pure condensate, there have also been a growing interest in finite-temperature properties~\cite{Erhard2004, Schmaljohann2004,Yuki2008}. In our previous paper, we discussed the spin-wave collective modes with dipole symmetry above ${T_{\rm BEC}}$~\cite{Yuki2008}. We showed that the spin-1 Bose gas exhibits a fascinating dynamics even if above ${T_{\rm BEC}}$ and we indicated that the spin-1 Bose condensate gas below ${T_{\rm BEC}}$ will exhibit much richer dynamics due to the interaction between condensate and noncondensate atoms in addition to the spin-spin interaction.

One of the most interesting experiments of the dynamics of spin-1 Bose condensed gases at finite temperatures is the population dynamics of each internal state~\cite{Erhard2004}. In this experiment, one found that the condensates do not conserve the total spin by themselves. In addition, the existence of the condensate enhances equilibration of the noncondensate component~\cite{Nikuni2001}. We are interested in this property.

In this paper, we first derive rate equations for spin-1 Bose gases above ${T_{\rm BEC}}$. Solving the rate equations, we discuss the population dynamics. We estimate characteristic time required to reach equilibrium. In the experiment, the time evolution of the particle numbers of each internal state was observed below ${T_{\rm BEC}}$ where both condensate and noncondensate atoms are present~\cite{ Erhard2004}. Comparing our theory, which only includes noncondensate atoms, with this experiment, we discuss how the existence of the condensate atoms will affect to the population dynamics.

\section{SPIN-1 KINETIC EQUATION}
We consider a gas of bosonic atoms with spin $F=1$ in an optical trap, so called spin-1 Bose gas. Each atom has three hyper-fine spin states; ${\it m_F}{=1,0,-1}$. In second quantized form, the Hamiltonian for this system is given by
\begin{eqnarray}
\hat{H} &=& \sum_{ij} \int d\textbf{r} \hat \Psi _i^\dagger \left( \textbf{r},t \right) \langle i \mid \uuline{H_0} \left( \textbf{r} \right) \mid j \rangle \hat \Psi _j  \left( \textbf{r},t \right) \notag \\
     & & +\frac{g_0}{2}\sum_{i,j}\int d \textbf{r} \hat \Psi _i^\dagger  \left( \textbf{r},t \right) \hat \Psi _j^\dagger  \left( \textbf{r},t \right) \hat \Psi _j  \left( \textbf{r},t \right) \hat \Psi _i  \left( \textbf{r},t \right) \notag \\
& & +\frac{g_2}{2} \sum_{i,j,i^\prime , j^\prime }\sum_{\alpha } \int d \textbf{r} \hat \Psi _i^\dagger  \left( \textbf{r},t \right) \hat \Psi _{i^\prime }^\dagger  \left( \textbf{r},t \right) S_{ij}^\alpha S_{i^\prime j^\prime } ^\alpha \hat \Psi _{j^\prime } \left( \textbf{r},t \right) \hat \Psi _j \left( \textbf{r},t \right), \label{H}
\end{eqnarray}
where $\hat \Psi_i^\dagger \left( \textbf{r},t \right)$ is the Bose field operator of the Heisenberg expression satisfying the equal-time commutation relation: $
\left[ \hat \Psi_i \left( \textbf{r},t \right), \hat\Psi_j^\dagger \left( \textbf{r}^\prime,t \right) \right] = \delta_{ij} \delta \left( \textbf{r} -\textbf{r}^\prime \right)$,
and $i, j$ indicate the hyperfine spin states $m_F$. The hat indicates a second quantized operator. Here, $\uuline{H_0}\left(\textbf{r},t\right)$ is the single atom Hamiltonian given by
\begin{eqnarray}
\uuline{H_0} \left( \textbf{r},t \right) &=& \left[ - \frac{\hbar^2}{2m}{\nabla}_{\textbf{r}}^2 + V \left( \textbf{r}\right) \right] \uuline{1} \nonumber \\
	& & +g\mu_B \sum_{\alpha}B^{\alpha}\left( \textbf{r},t \right) \uuline{S}^{\alpha} + \sum_{\alpha\beta}B_q^{\alpha\beta}\left( \textbf{r},t \right) \uuline{Q}^{\alpha\beta},\label{H_0}
\end{eqnarray}
where the double underline indicates a $3\times 3$ matrix, and $\alpha,\beta$ indicate spin components; $x, y, z$. The first term of Eq.(\ref{H_0}) describes the center-of-mass motion of an atom in a harmonic trap $V \left( \textbf{r} \right)= \frac{m}{2} \left( \omega_x^2x^2 + \omega_y^2y^2 + \omega_z^2z^2 \right)$, where $m$ is the mass of a single atom. The second term describes the linear Zeeman energy, where $B^{\alpha}$ is spin-$\alpha$ component of the magnetic field, and $\uuline{S}^{\alpha} \left( \alpha = x,y,z \right)$ denotes the spin-1 matrix of spin-$\alpha$ component, and $g$ is Lande $g$ factor and $\mu_B$ is Bohr magneton. The third term is the quadratic Zeeman energy, in which we introduced the quadratic magnetic field $B_q^{\alpha\beta}$ and the quadrupole operator $\uuline{Q}^{\alpha\beta}$ is given by $\uuline{Q}^{\alpha \beta }= \left( 1-\frac{1}{2} \delta _{\alpha \beta }  \right) \left( \uuline{S}^\alpha \uuline{S}^\beta + \uuline{S}^\beta \uuline{S}^\alpha - \delta _{\alpha \beta }\right)\frac{2}{3} \uuline{S}^2 $. The coupling constants $g_0$ and $g_2$ are given by $g_0=\left(4\pi\hbar^2/3m\right)\left(a_0+2a_2\right)$ and $ g_2=\left(4\pi\hbar^2/3m\right)\left(a_2-a_0\right)$, $a_0$ and $a_2$ being the {\it s}-wave scattering lengths for collision channels with total spins ${\cal F}=0$ and $2$, respectively.  In this system, only coupling process between states with $m_F=0$ and $m_F=\pm 1$ can be permitted.

The time evolution of the nonequilibrium system is described by the Wigner distribution function
\begin{eqnarray}
{W}_{ij} \left( \textbf{p},\textbf{r} \right) = \int d \textbf{r}^\prime e^{i\textbf{p} \cdot \textbf{r}^\prime / \hbar }%
          \langle\hat \Psi ^\dagger _{j}\left( \textbf{r} + \textbf{r}^\prime /2 \right)\hat \Psi _{i}\left( \textbf{r} - \textbf{r}^\prime /2 \right)\rangle,
\end{eqnarray}
Knowledge of this function allows one to calculate various nonequilibrium physical quantities, such as the local density given by
\begin{eqnarray}
\uuline{n} \left( \textbf{r},t \right) = \int \frac{d \textbf{p}}{\left( 2\pi \hbar \right)^3} \uuline{W} \left( \textbf{p},\textbf{r},t \right),
\end{eqnarray}
where $n_{ij} \left( \textbf{r},t \right) \equiv \langle \hat \Psi_i^\dagger \left( \textbf{r},t \right) \hat \Psi_j \left( \textbf{r},t \right) \rangle$. Using a semiclassical approximation to describe atomic motion in terms of a phase-space distribution function, we obtain the semiclassical kinetic equation \cite{Yuki2008}: 
\begin{eqnarray}
\frac{\partial W_{ij}}{\partial t} +\frac{\textbf{p}}{m} \cdot \nabla _{\textbf{r}} W_{ij} - \frac{1}{2} \left\{ \nabla _{\textbf{r}} \uuline{U},\nabla _{\textbf{p}} \uuline{W} \right\} _{ij}+ \frac{i}{\hbar} \left[ \uuline{U} , \uuline{ W}\right]_{ij}=I_{ij}, \ \ \ \ \label{Wigner100}
\end{eqnarray}
where  the effective potential is described as
\begin{eqnarray}
\uuline{U}\equiv V\uuline{1}+g_0  {n}\uuline{1}+g_0\uuline{ {n}}+\sum_{\alpha}\left\{g\mu_BB^\alpha\uuline{S}^\alpha+g_2 {M}^\alpha\uuline{S}^\alpha+g_2\uuline{S}^\alpha\uuline{ {n}}\uuline{S}^\alpha\right\}+\sum_{\alpha\beta}B_q^{\alpha\beta}\uuline{Q}^{\alpha\beta}.\label{G_Un}
\end{eqnarray}
We defined the magnetization as
$ M^\alpha\left(\textbf{r},t \right)\equiv{\rm Tr}\left[\uuline{n}\left(\textbf{r},t\right)\uuline{S}^\alpha\right]$. The collision integral $I_{ij}$ on the right hand side of Eq.(\ref{Wigner100}) describes collisions between atoms, whose explicit expression is given in Ref.\cite{Yuki2008}

Throughout this paper we assume relatively high temperature regime where the phase space density is sufficiently low, i.e. $W_{ij}\ll 1$. Under this assumption, in the collision integral $I_{ij}$, we only retain terms to second order in $W_{ij}$. This is equivalent to neglecting quantum degeneracy in the center of mass motion of atoms.

The static thermal equilibrium distribution is determined from the condition $I_{ij}=0$. This leads to the Maxwell-Boltzmann distribution
\begin{eqnarray}
W_{ij}^0=\delta_{ij}W_{ii}^0= e^{-\beta_i\left(\epsilon_i-\mu_i\right)}, \label{kinnzi_2_3}
\end{eqnarray}
where $\epsilon_i$ is the single-particle excitation energy 
\begin{eqnarray}
{\epsilon}_i\equiv\frac{p^2}{2m}+{U}_{ii},
\end{eqnarray}
and ${\mu}_i$ is the chemical potential of the {\it i} state. The condition for the equilibrium chemical potentials is given by
\begin{eqnarray}
 {\mu}_1+ {\mu}_{-1}=2 {\mu}_0\label{zyoukenn}.
\end{eqnarray}

\section{Rate Equations for the population dynamics above the Bose-Einstein transition temperature}
We consider the time evolution of atoms in each internal state above the transition temperature. The particle number of the internal state $i$ is given by 
\begin{eqnarray}
N_i\left(t\right)=\int d\textbf{r}\int \frac{d\textbf{p}}{\left(2\pi\hbar\right)^3}W_{ii}\left(\textbf{p},\textbf{r},t\right).\label{N_keisann}
\end{eqnarray}
The equation for the particle number is obtained by integrating kinetic equation Eq.(\ref{Wigner100}) over $\textbf{r}$ and $\textbf{p}$. In this section, we assume that the system is close to equilibrium at each time, and thus approximate the distribution function with the Maxwell-Boltzmann form:
\begin{eqnarray}
W_{ij}\left(\textbf{p},\textbf{r},t\right)=\delta_{ij}f_i\left(\textbf{p},\textbf{r},t\right)=\delta_{ij}\exp\left\{ -\left[ \epsilon_i\left(\textbf{p},\textbf{r},t\right)-\mu_i\left(t\right) \right] / k_BT_i\left(t\right)\right\}
, \label{kinnzi_22_3}
\end{eqnarray}
where $\mu_i$ and $T_i$ depend on time. With this approximation, the collision integral reduces to 
\begin{eqnarray}
I_{ii}&=&\frac{\pi}{\hbar}\sum_{ji^\prime j^\prime }\sum_{\alpha\beta}\int\frac{d\textbf{p}_2}{\left(2\pi\hbar\right)^3}\int\frac{d\textbf{p}_3}{\left(2\pi\hbar\right)^3}\int d\textbf{p}_4\delta\left(\textbf{p}+\textbf{p}_2-\textbf{p}_3-\textbf{p}_4\right)\delta\left(\epsilon_{ip}+\epsilon_{jp_2}-\epsilon_{i^\prime p_3}-\epsilon_{j^\prime p_4}\right)\nonumber \\
	& & \times 2g_2^2\left(S^\alpha_{ii^\prime}S^\beta_{i^\prime i}S^\alpha_{jj^\prime}S^\beta_{j^\prime j}+S^\alpha_{ij^\prime}S^\beta_{j^\prime j}S^\alpha_{ji^\prime}S^\beta_{i^\prime i}\right) \left[f_{i^\prime}\left(\textbf{p}_3\right)f_{j^\prime}\left(\textbf{p}_4\right)-f_{i}\left(\textbf{p}\right)f_{j}\left(\textbf{p}_2\right)\right] .\label{G_G_G}
\end{eqnarray}
We note that the terms proportional to $g_0^2$ and $g_0g_2$ in the collision integral vanish because of conservation of momentum and energy.

Next, we integrate Eq.(\ref{Wigner100}) over $\textbf{r}$ and $\textbf{p}$ to obtain rate equations for the particle number. For simplicity, we neglect the Hartree-Fock effective potential in the single-particle energy $\epsilon_i\left(\textbf{p},\textbf{r},t\right)$, i.e.
\begin{eqnarray}
\epsilon_i\left(\textbf{p},\textbf{r},t\right)\approx\epsilon\left(\textbf{p},\textbf{r}\right)=\frac{p^2}{2m}+V .\label{kinnzi_3_1}
\end{eqnarray}
We also assume that the temperatures of three internal states are close to each other, so we only retain terms to second order in fluctuations of $\beta_{i}$.

All the terms of the left side of Eq.(\ref{Wigner100}) other than the first term do not contribute because those integrand are odd functions. Thus, we have only to integrate Eq.(\ref{G_G_G})
\begin{eqnarray}
\frac{d N_i}{d t}=\int d\textbf{r}\int \frac{d\textbf{p}}{\left(2\pi\hbar\right)^3}I_{ii}\left(\textbf{p},\textbf{r},t\right).\label{dN_dt}
\end{eqnarray}

Using (\ref{kinnzi_22_3}) and (\ref{G_G_G}) in (\ref{dN_dt}) and performing integrals explicitly, we obtain the equation of motion for particle number as
\begin{eqnarray}
\frac{d N_i}{dt}&=&2\frac{\gamma_i}{\beta_i}\sum_{ji^\prime j^\prime}\sum_{\alpha\beta}\left(S_{ii^\prime}^\alpha S_{i^\prime i}^\beta S_{jj^\prime}^\alpha S_{j^\prime j}^\beta+S_{ij^\prime}^\alpha S_{j^\prime j}^\beta S_{ji^\prime}^\alpha S_{i^\prime i}^\beta\right)\nonumber \\
	& & \ \ \ \ \ \ \ \ \ \ \times\left\{\frac{\beta_{i^\prime}^3\beta_{j^\prime}^3}{\left[\left(\beta_{i^\prime}+\beta_{j^\prime}\right)/2\right]^5}N_{i^\prime}N_{j^\prime}-\frac{\beta_{i}^3\beta_{j}^3}{\left[\left(\beta_{i}+\beta_{j}\right)/2\right]^5}N_{i}N_{j}\right\}.\label{G_N_2}
\end{eqnarray}
Here, we defined the collisional relaxation rate associated with the population transfer as
\begin{eqnarray}
\gamma_i\equiv g_2^2\frac{\beta_i}{\hbar}\frac{\left(m\omega_{ho}\right)^3}{\left(2\pi\hbar\right)^3},\label{G_gamma}
\end{eqnarray}
where $\omega_{ho}\equiv\left(\omega_x\omega_y\omega_z\right)^{1/3}$. We write down explicit forms of Eq.(\ref{G_N_2}) for three components
\begin{eqnarray}
\frac{d N_1}{d t}&=&4{\gamma_1}\left\{\frac{\beta_0}{\beta_1}N_0N_0-\frac{\beta_1^2\beta_{-1}^3}{\left[\left(\beta_1+\beta_{-1}\right)/2\right]^5}N_1N_{-1}\right\},\label{G_N_1}\\
\frac{d N_0}{d t}&=&8{\gamma_0}\left\{ \frac{\beta_1^3\beta_{-1}^3}{\beta_0\left[\left(\beta_1+\beta_{-1}\right)/2\right]^5}N_1N_{-1}-N_0N_0 \right\},\label{G_N_0}\\
\frac{d N_{-1}}{d t}&=&4\gamma_{-1}\left\{\frac{\beta_0}{\beta_{-1}}N_0N_0-\frac{\beta_1^3\beta_{-1}^2}{\left[\left(\beta_1+\beta_{-1}\right)/2\right]^5}N_1N_{-1}\right\}.\label{G_N_-1}
\end{eqnarray}

From Eq.(\ref{G_N_2}), we find that population dynamics depends not only on the particle numbers but also on the temperatures. In general, the atoms flow into the state with lower temperature. In addition, depending on the ratios of the each particle number, spin-spin interaction ($g_2$) contributes to increasing $m_F=0$ particle, or vice versa.

\begin{figure}
  \begin{center}
   \scalebox{0.7}[0.7]{\includegraphics{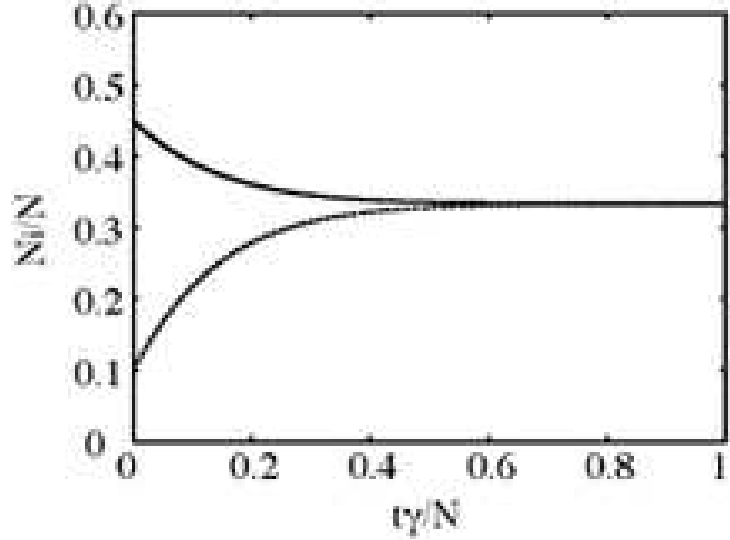}}
    \caption{Time evolution of spin populations in a trapped gas of ${\rm ^{87}Rb}$ atoms. We take the following values for the parameters: $N =1\times 10^5$, $\left\{\omega_x/2\pi,\omega_y/2\pi,\omega_z/2\pi\right\}=\left\{890,890,160\right\} {\rm Hz}$ and $T=500{\rm nK}$. We set the initial condition as $N_1:N_0:N_{-1}=45:10:45$. With this initial condition, the time evolutions of $N_1$ obtained from Eq.(\ref{G_N_1}) and $N_{-1}$ obtained from Eq.(\ref{G_N_-1}) are precisely the same, and are indicated by the single solid line. The broken line indicates the time evolution of $N_0$ obtained from Eq.(\ref{G_N_0}).}
    \label{fig:N_picture1}
  \end{center}
\end{figure}

In general, temperatures vary from time to time. So, we will also derive the equations of motion for the temperatures. First, we derive the equation for the energy with neglecting the mean field. The energy of the internal state $i$ is given by 
\begin{eqnarray}
E_i\left(t\right)=\int d\textbf{r}\int\frac{d\textbf{p}}{\left(2\pi\hbar\right)^3}\epsilon_i\left(\textbf{p},\textbf{r}\right)f_i\left(\textbf{p},\textbf{r},t\right).
\end{eqnarray}
From the above equation, we can derive the equation for the energy by integrating the kinetic equation (\ref{Wigner100}) multiplied by the single-particle energy $\epsilon_i$ over real and momentum space. Using the approximations (\ref{kinnzi_22_3}) and (\ref{kinnzi_3_1}), we obtain the expression for the energy in terms of $\beta_i$ and $\mu_i$
\begin{eqnarray}
E_i&\approx&\frac{3}{\beta_i}\frac{1}{\left(\beta_i\hbar\omega_{ho}\right)^3} e^{\beta_i\mu_i}.
\end{eqnarray}
The energy $E_i$ can also be written in terms of the particle $N_i$ and the temperature parameter $\beta_i$ as
\begin{eqnarray}
E_i=\frac{3N_i}{\beta_i}.\label{E_i_b_i}
\end{eqnarray}
With the assumption that the Wigner function obeys the Boltzmann distribution function (\ref{kinnzi_22_3}), we obtain the equation for the energy as
\begin{eqnarray}
\frac{d E_i}{d t}=\int d\textbf{r}\int \frac{d\textbf{p}}{\left(2\pi\hbar\right)^3}\epsilon_i\left(\textbf{p},\textbf{r}\right)I_{ii}\left(\textbf{p},\textbf{r},t\right).\label{118}
\end{eqnarray}

Using (\ref{kinnzi_22_3}) and (\ref{G_G_G}) in (\ref{118}) and performing integrals explicitly, we obtain the equation of motion for the energy as
\begin{eqnarray}
\frac{d E_i}{d t}&=&2g_2^2\frac{\left(m\omega_{ho}\right)^3}{\hbar\left(2\pi\hbar\right)^{3}}\sum_{ji^\prime j^\prime}\sum_{\alpha\beta}\left(S_{ii^\prime}^\alpha S_{i^\prime i}^\beta S_{jj^\prime}^\alpha S_{j^\prime j}^\beta+S_{ij^\prime}^\alpha S_{j^\prime j}^\beta S_{ji^\prime}^\alpha S_{i^\prime i}^\beta\right)\nonumber \\
	 & & \ \ \ \ \ \ \ \ \ \ \ \ \ \ \ \ \ \ \ \ \ \ \times\Biggl(\left\{\frac{5}{18}\frac{\beta_{i^\prime}^4\beta_{j^\prime}^4}{\left[\left(\beta_{i^\prime}+\beta_{j^\prime}\right)/2\right]^6}-\frac{1}{9}\frac{\beta_{i^\prime}^4\beta_{j^\prime}^4\left(\beta_{i^\prime}-\beta_{j^\prime}\right)}{\left[\left(\beta_{i^\prime}+\beta_{j^\prime}\right)/2\right]^7}\right\}E_{i^\prime}E_{j^\prime}\Biggr.\nonumber \\
	 & & \ \ \ \ \ \ \ \ \ \ \ \ \ \ \ \ \ \ \ \ \ \ \ \ -\Biggl. \left\{\frac{5}{18}\frac{\beta_i^4\beta_j^4}{\left[\left(\beta_{i}+\beta_{j}\right)/2\right]^6}-\frac{1}{9}\frac{\beta_i^4\beta_j^4\left(\beta_{i}-\beta_{j}\right)}{\left[\left(\beta_{i}+\beta_{j}\right)/2\right]^7}\right\}E_{i}E_{j}  \Biggr).\label{G_E_2}
\end{eqnarray}

Using (\ref{E_i_b_i}) and combing (\ref{G_N_2}) and (\ref{G_E_2}), we obtain the equations of motion for the temperatures. 
\begin{eqnarray}
\frac{d\beta_i}{dt}&=&2g_2^2\frac{\left(m\omega_{ho}\right)^3}{\hbar\left(2\pi\hbar\right)^3}\sum_{ji^\prime j^\prime}\sum_{\alpha\beta}\left(S_{ii^\prime}^\alpha S_{i^\prime i}^\beta S_{jj^\prime}^\alpha S_{j^\prime j}^\beta+S_{ij^\prime}^\alpha S_{j^\prime j}^\beta S_{ji^\prime}^\alpha S_{i^\prime i}^\beta\right)\nonumber \\
	& & \ \ \ \ \ \ \ \ \ \ \times\Biggl(\frac{\beta_{i^\prime}^3\beta_{j^\prime}^3}{\left[\left(\beta_{i^\prime}+\beta_{j^\prime}\right)/2\right]^5}\frac{\beta_iN_{i^\prime}N_{j^\prime}}{N_i}\left\{1-\frac{5}{6}\frac{\beta_i}{\left[\left(\beta_{i^\prime}+\beta_{j^\prime}\right)/2\right]}+\frac{1}{3}\frac{\beta_i\left(\beta_{i^\prime}-\beta_{j^\prime}\right)}{\left[\left(\beta_{i^\prime}+\beta_{j^\prime}\right)/2\right]^2}\right\}\Biggr.\nonumber \\
	& & \ \ \ \ \ \ \ \ \ \ \ \ \ -\Biggl.\frac{\beta_{i}^4\beta_{j}^3}{\left[\left(\beta_{i}+\beta_{j}\right)/2\right]^5}N_{j} \left\{1-\frac{5}{6}\frac{\beta_i}{\left[\left(\beta_{i}+\beta_{j}\right)/2\right]}-\frac{1}{3}\frac{\beta_i\left(\beta_{i}-\beta_{j}\right)}{\left[\left(\beta_{i}+\beta_{j}\right)/2\right]^2}\right\} \Biggr).\label{beta_kinetic_2}
\end{eqnarray}

We now consider the special situation where the temperatures of three internal states are the same. In this case, the temperature stays constant in time. In Fig.\ref{fig:N_picture1}, we plot the population dynamics in the situation where all the temperatures remain the same constant value. We plot the variation of the populations for ${\rm ^{87}Rb}$, where the total number of atoms is $N=1\times 10^5$ and the temperature is $T=500{\rm nK}$, which corresponds to $T\approx 2T_{\rm {BEC}}$. We referred several quantities to the experiment of Ref.\cite{Erhard2004}: the trap frequencies are $\omega_z/2\pi=160 {\rm Hz}$ and $\omega_{\perp}=890 {\rm Hz}$, and the initial situation is $N_1:N_0:N_{-1}=45:10:45$. In Fig.\ref{fig:N_picture1}, we find that population numbers evolve toward the situation where the numbers of particles are the same, where the chemical potentials satisfy Eq.(\ref{zyoukenn}). This is consistent with the behavior of the noncondensate atoms observed in the experiment of Ref.\cite{Erhard2004}.

Next, we estimate characteristic time required for the system to reach equilibrium. From Eqs.(\ref{G_N_2}) and (\ref{G_gamma}), we see that the characteristic time for the variation of the population is $1/\gamma_i$. The relaxation rate $\gamma_i$ is very small because $\gamma_i\propto g_2^2$, where $g_2$ is the spin-spin interaction ($\left| g_0 \right| \approx 200\left| g_2 \right|$ for ${\rm ^{87}Rb}$) as shown in Eq.(\ref{G_gamma}). A rough estimate gives the relaxation time as $1/\gamma_i\times N_{\it total}\simeq 39 {\rm s}$. From the result in Fig.\ref{fig:N_picture1} for $T=500{\rm nK}$, we determine the characteristic relaxation time ${\it t}$ as the time at which $N_i$ reach equilibrium value. We find $t\approx 27{\rm s}$. This turns out to be the order of the lifetime of trapped atoms. We find that the populations change rapidly in early time when they are far from equilibrium. The higher the ratio of each particle number of the internal state, the more rapid the particle numbers vary. In the experiment of Ref.\cite{Erhard2004}, the time required for populations to reach equilibrium is $t\approx 10{\rm s}$. This is less than half of our result for a thermal gas. Since in the experiment, both the condensate and noncondensate atoms exist, we speculate that the interaction between the condensate and noncondensate atoms equilibrates the system much more rapidly~\cite{Nikuni2001}. We note that the relaxation rate (\ref{G_gamma}) can be written in terms of the density in the center of the trap potential $n_i\left(0\right)$:
\begin{eqnarray}
\gamma_i=\frac{g_2^2m^2}{8\pi\hbar^4}\sqrt{\frac{8}{\pi m\beta_i}}\frac{n_i\left(0\right)}{N_i}.
\end{eqnarray}
In the case of a trapped single-component Bose gas, Ref.~\cite{Nikuni2001} showed that in the BEC phase, condensate density makes additional contribution to the relaxation rate. Thus, in a trapped gas, the collisional relaxation time become much shorter in the BEC phase because of the high condensate density at the center of the trap.

\section{conclusions}
In this paper, we studied the population dynamics of the spin-1 Bose atoms above ${T_{\rm BEC}}$
. We found that the time evolution depend on both relative number and relative temperature. Comparing our result with the experiment below ${T_{\rm BEC}}$, where the condensate is present~\cite{Erhard2004}, we found that in the case where both the condensate and noncondensate atoms exist, equilibrium is reached in less than half the time it takes when condensate atoms are absent. This suggests that the condensate atoms help the equilibration of the noncondensate atoms. We also found that the initial time evolution of the population number is very rapid nevertheless its relaxation time is very large.

In future work, we will derive a kinetic theory for spin-1 Bose condensed gases below ${T_{\rm BEC}}$, where the  condensate and the noncondensate atoms interact with each other. This theory will consist of coupled equations for the condensate and noncondensate. Using this theory, we will study the population dynamics at finite temperatures below ${T_{\rm BEC}}$, where both condensate and the noncondensate atoms are present. We will also consider coupled collective motion of the condensate and noncondensate.


\begin{thebibliography}{10}

\bibitem{Anderson1995}
M.~H. Anderson, J.~R. Ensher, M.~R. Matthews, C.~E. Wieman, and E.~A. Cornell,
  Science {\bf 269},  198  (1995).

\bibitem{Davis1995}
K.~B. Davis, M.~O. Mewes, M.~R. Andrews, N.~J. van Druten, D.~S. Durfee, D.~M.
  Kurn, and W. Kitterle, Phys. Rev. Lett. {\bf 75},  3969  (1995).

\bibitem{Myatt1997}
C.~J. Myatt, E.~A. Burt, R.~W. Ghrist, E.~A. Cornell, and C.~E. Wieman, Phys.
  Rev. Lett. {\bf 78},  586  (1997).

\bibitem{Hall1998}
D.~S. Hall, M.~R. Matthews, J.~R. Ensher, C.~E. Wieman, and E.~A. Cornell,
  Phys. Rev. Lett. {\bf 81},  1539  (1998).

\bibitem{Lewandowski2002}
H.~J. Lewandowski, D.~M. Harber, D.~L. Whitaker, and E.~A. Cornell, Phys. Rev.
  Lett. {\bf 88},  070403  (2002).

\bibitem{McGuirk2002}
J.~M. McGuirk, H.~J. Lewandowski, D.~M. Harber, T. Nikuni, J.~E. Williams, and
  E.~A. Cornell, Phys. Rev. Lett. {\bf 89},  090402  (2002).

\bibitem{Oktel2002_2}
M.~{\"O}. Oktel and L.~S. Levitov, Phys. Rev. Lett. {\bf 88},  230403  (2002).

\bibitem{Fuchs2002}
J.~N. Fuchs, D.~M. Gangardt, and F. Lalo{\"e}, Phys. Rev. Lett. {\bf 88},
  230404  (2002).

\bibitem{Williams2002}
J.~E. Williams, T. Nikuni, and Charles~W. Clark, Phys. Rev. Lett. {\bf 88},
  230405  (2002).

\bibitem{Nikuni2002}
T. Nikuni, J.~E. Williams, and C.~W. Clark, Phys. Rev. A {\bf 66},  043411
  (2002).

\bibitem{Nikuni2003}
T. Nikuni and J.~E. Williams, J. Low. Temp. Phys. {\bf 133},  323  (2003).

\bibitem{Stenger1998}
J. Stenger, S. Inouye, D.~M. Stamper-Kurn, H.~J. Miesner, A.~P. Chikkatur, and
  W. Ketterle, Nature {\bf 396},  345  (1998).

\bibitem{Kurn1998}
D.~M. Stamper-Kurn, M.~R. Andrews, A.~P. Chikkatur, S. Inouye, H.~J. Miesner,
  J. Stenger, and W. Ketterle, Phys. Rev. Lett. {\bf 80},  2027  (1998).

\bibitem{Law1998}
C.~K. Law, H. Pu, and N.~P. Bigelow, Phys. Rev. Lett. {\bf 81},  5257  (1998).

\bibitem{Miesner1999}
H.~J. Miesner, D.~M. Stamper-Kurn, J. Stenger, S. Inouye, A.~P. Chikkatur, and
  W. Ketterle, Phys. Rev. Lett. {\bf 82},  2228  (1999).

\bibitem{Zhang2003}
Wenxian Zhang, Su Yi, and Li You, New J. Phys. {\bf 5},  77  (2003).

\bibitem{Chang2004}
M.~S. Chang, C.~D. Hamley, M.~D. Barrett, J.~A. Sauer, K.~M. Fortier, W. Zhang,
  L. You, and M.~S. Chapman, Phys. Rev. Lett. {\bf 92},  140403  (2004).

\bibitem{Schmaljohann2004}
H. Schmaljohann, M. Erhard, J. Kronj{\"a}ger, K. Sengstock, and K. Bongs, Appl.
  Phys. B {\bf 79},  1001   (2004).

\bibitem{Erhard2004}
M. Erhard, H. Schmaljohann, J. Kronj{\"a}ger, K. Bongs, and K. Sengstock, Phys.
  Rev. A {\bf 70},  031602R  (2004).

\bibitem{Higbie2005}
J.~M. Higbie, L.~E. Sadler, S. Inouye, A.~P. Chikkatur, S.~R. Leslie, K.~L.
  Moore, V. Savalli, and D.~M. Stamper-Kurn, Phys. Rev. Lett. {\bf 95},  050401
   (2005).

\bibitem{Sadler2006}
L.~E. Sadler, J.~M. Higbie, S.~R. Leslie, M. Vengalattore, and D.~M.
  Stamper-Kurn, Nature {\bf 443},  312  (2006).

\bibitem{Yuki2008}
Yuki Endo and Tetsuro Nikuni, J. Low. Temp. Phys. {\bf 152},  21  (2008).

\bibitem{Nikuni2001}
T. Nikuni and A. Griffin, Phys. Rev. A {\bf 65},  011601  (2001).

\end{thebibliography}

\end{document}